\documentclass[aps,prx,groupedaddress,twocolumn,showpacs,longbibliography,10pt]{revtex4-2}
\usepackage{amsmath}
\usepackage{amsthm, amssymb,bbold,slashed}
\usepackage{braket}
\usepackage{graphicx}
\usepackage{draft}
\usepackage{color}
\usepackage{times, verbatim}
\usepackage{mathtools,tikz}
\usetikzlibrary{matrix}
\usepackage{natbib}
\usepackage{multirow}
\usepackage{hyperref}
\usepackage{subcaption}

\definecolor{colour1}{rgb}{0.368417, 0.506779, 0.709798}
\definecolor{colour2}{rgb}{0.880722, 0.611041, 0.142051}
\definecolor{colour3}{rgb}{0,1,1}
\definecolor{colour4}{rgb}{0,1,0}
\definecolor{colour5}{rgb}{1,1,0}

\usepackage{amsthm}

\hypersetup{
  colorlinks=true,
}

\makeatletter
\newsavebox{\@brx}
\newcommand{\llangle}[1][]{\savebox{\@brx}{\(\m@th{#1\langle}\)}%
\mathopen{\copy\@brx\kern-0.5\wd\@brx\usebox{\@brx}}}
\newcommand{\rrangle}[1][]{\savebox{\@brx}{\(\m@th{#1\rangle}\)}%
\mathclose{\copy\@brx\kern-0.5\wd\@brx\usebox{\@brx}}}
\makeatother

\begin{document}

\author{Fedor K.~Popov}

\affiliation{Simons Center for Geometry and Physics, SUNY, Stony Brook, NY 11794, USA}

\title{
  On Real Time Dynamics of Large $N$ Models
}

\begin{abstract}
  We analyze the real-time dynamics of the large
  $N$ vector model, focusing on heavy states with energies of the order  $N$.
  In this regime, we demonstrate that interactions become sufficiently strong to produce non-zero condensate of the Hubbard-Stratonovich field
  $\sigma$, which, in turn, induces particle production. This process leads to a significant transformation of the initial state and potential thermalization.
  For homogeneous perturbations, our results show that the equations become integrable, yet can still lead to thermalization in the continuum limit.
  Furthermore, we calculate the energies of these heavy states and their contributions to the thermal free energy, thereby determining the free energy of the critical
  $O(N)$ model by operator counting.
\end{abstract}

\date{\today}


\maketitle

\section{Introduction}
The solvable models of quantum field theories play a crucial role in the development of new ideas, enhancing our understanding of various aspects of other quantum field theories in similar setups. In particular, certain models become solvable when the number of fields $N$ is taken to infinity \cite{Klebanov:2018fzb, Wilson:1973jj, Moshe:2003xn}. Examples include large $N$ vector models, which have been extensively studied due to their simplicity and solvability. These models are believed to be integrable in the large $N$ limit, thanks to the presence of approximately conserved higher spin currents \cite{Maldacena:2011jn}. Additionally, the coupling constant scales as $\frac{1}{N}$, and some correlators involving light operators receive negligible contributions in the large $N$ limit. Consequently, it is commonly believed that such models cannot thermalize at infinite $N$ \cite{Chowdhury:2017jzb, milekhin2023all}. However, this logic works only for low excited states, and to make a precise statement about thermalization, it is essential to study the behavior of highly excited states, as we expect that at temperatures $\beta \sim 1$, the relevant energy scale should be on the order of $N$, i.e., $E \sim N E_0$\footnote{An analogous statement was made for large $N$ matrix models but formulated as a constraint in terms of temperatures \cite{Shenker:2011zf}}.

In this paper, we argue that high-energy states in large $N$ theories, contrary to common belief, could thermalize due to non-perturbative effects involving particle production. The collective real-time dynamics of the fields in these models create a non-stationary background described by the Hubbard-Stratonovich field $\sigma$, which subsequently leads to particle creation and possible thermalization. Although these equations are analytically challenging, they can be studied numerically. In the case of homogeneous excitations, our numerical studies indicate that a system with a finite number of degrees of freedom does not thermalize and, moreover, is actually integrable. Thus, we find a set of commuting charges that are conserved. Nonetheless, in the thermodynamic or continuum limit, when we have an infinite number of degrees of freedom, these systems could still lead to thermalization. By examining the stationary solutions of these equations, we also discovered a new family of states that generalizes those found by Giombi et al. \cite{Giombi:2020enj}. By counting these states, we determine the thermal free energy. For the critical $O(N)$ model, we obtain a result that coincides with the one obtained using conventional techniques \cite{chubukov1994theory}.

Another interesting application of this approach is studying the real-time dynamics of such models in the background of non-trivial electromagnetic or gravitational backgrounds, such as constant electric fields or gravitational de Sitter backgrounds. This approach could potentially enhance our understanding of quantum field theories in these settings. Addressing these questions is crucial, as the behavior of these theories in such backgrounds — whether they reach a steady or equilibrium state or create a strong backreaction that significantly alters the background — remains unclear. These questions will be addressed in future papers.

The paper is organized as follows. In Section \ref{sec1}, we briefly review the Schwinger-Keldysh technique, which describes the real-time dynamics of quantum field theories,
and apply it to the large $N$ vector models. In Section \ref{sec2}, we derive the equations governing the dynamics of the large $N$ quantum vector model in the regime involving high-energy states,
demonstrating that such states create a non-zero $\sigma$ field background, which subsequently leads to particle production and introduces significant corrections to the initial states.
We also show that these equations, in the homogeneous setup, are integrable and present the set of commuting charges. In Section \ref{sec3}, we count the heavy states and compute their energy spectrum, which allows us to determine the thermal free energy of the large $N$ models at temperature $\beta$.

\section{Review of the large $N$ vector model and Schwinger-Keldysh technique}
\label{sec1}
In this section, we review the large $N$ models and the Schwinger-Keldysh technique \cite{Moshe:2003xn, arseev2015nonequilibrium}. We focus on the three-dimensional case, $d=3$, as it simplifies some aspects of the analysis and avoids renormalization issues encountered in $d=2$ and $d=4$. Furthermore, we assume that our space is $\mathcal{M}_3 = \mathbb{R} \times \mathcal{M}_2$, where the spatial manifold $\mathcal{M}_2$ is compact. The large $N$ model is described by the following action:
\begin{gather}
  S = \int d^3 x \left[\frac12 \left(\partial_\mu \phi_i\right)^2 + \frac14 R \phi^2
    - \frac12 m_0^2 \phi_i^2 - \frac{g_0 \Lambda}{4N} \left(\phi_i^2\right)^2 \right],
\end{gather}
The index $i=1, \ldots, N$ enumerates the fields, and the theory exhibits $O(N)$ invariance. We explicitly introduce the UV cutoff $\Lambda$ and bare parameters $m_0$ and $g_0$, which must be renormalized to ensure a well-defined quantum field theory, with the physical mass $m \ll \Lambda$. The term $\frac{1}{4} R \phi^2$ ensures that the stress-energy tensor is traceless. To ensure a well-behaved large $N$ limit, we scale the coupling constant in front of the quartic interaction as $\frac{1}{N}$ to compensate for the additional sum over indices. This model is typically solved using the Hubbard-Stratonovich transformation, which introduces the field $\sigma$ to decouple the quartic term, followed by the saddle-point approximation.

Although this method is quite effective, we will adopt a different approach by directly considering the Feynman diagrams and identifying the leading contributions that survive in the large $N$ limit. This approach is also extremely useful for studying the real-time dynamics of large $N$ vector models using the Schwinger-Keldysh technique.
\begin{figure}
  \centering
  \begin{tikzpicture}
    \draw (-1,-1)--(0,0)--(1,-1);
    \draw[dashed] (0,0)--(0,0.5);
    \draw[line width=2] (0,1) circle (0.5);
    \node[left] at (-1,-1) {$\pm$};
    \node[right] at (1,-1) {$\pm$};
    \node[left] at (0,0) {$\pm$};
  \end{tikzpicture}
  \caption{Only "snail" diagrams survive in the large $N$ limit of the large $N$ vector model. The thick line corresponds to the exact propagator, where additional "snails" have already been taken into account.}
  \label{fig:dom}
\end{figure}
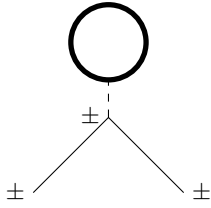

The Schwinger-Keldysh technique studies the real-time dynamics of various correlators by directly expanding the evolution operator and computing the resulting correlators for an arbitrary state $\ket{\Psi}$:
\begin{gather} \left\langle \Psi|U(-\infty, \infty) T\left[\phi_i(x_1,t_1) \phi_j(x_2,t_2) U(\infty,-\infty) \right) \right]| \Psi\rangle,\notag\\
  U(\infty,-\infty) = T\exp\left(-i \int\limits^\infty_{-\infty} H_I(t) , dt \right)\,. \label{eq:KStech} \end{gather} For stationary states, we typically omit the initial evolution operator $U$, but we will retain it here and simply expand both time-ordered exponentials. These correlators can be computed using the famous Wick theorem \cite{arseev2015nonequilibrium}. Now, instead of a single time-ordered correlator (the usual Feynman propagator), we have a matrix of correlators, as the fields can now belong to either the anti-time-ordered or time-ordered parts of the correlator. This leads to quite cumbersome expressions and significantly increases the number of diagrams. Let us show what these propagators look like in the case of free field theory. The operator of the field $\phi_i$ satisfies the following equation:
\begin{gather}
  \left(\partial_t^2 - \partial_i^2 + m^2\right) \phi_i =0, \\
  \left( - \partial_i^2 \right) h_{i\lambda}(x) = \lambda^2 h_\lambda(x),\, \sum_\lambda h_\lambda(x) h_\lambda^*(y) = \delta^{d-1}(x-y)  \notag
\end{gather}
then we can expand the field as
\begin{gather}
  \phi_i = \sum_\lambda \left[a_{i\lambda} f_{i\lambda}(x,t)  + {\rm h.c.} \right], \quad f_{i \lambda}(x,t)  = h_\lambda(x) f_{i \lambda}(t) \notag\\ [a_{i\lambda}, a^\dagger_{j\mu}] = \delta_{\lambda,\mu} \delta_{ij},\notag
\end{gather}
the operators $a_{i\mu}, a^\dagger_{j\nu}$ are our Fock operators and span Hilbert space.

Taking into account the usual commutation relations for fields, we find that the harmonics $f_\lambda(t)$ should behave as
\begin{gather}
  \forall \lambda \quad  i \left(f_\lambda\partial_t f^*_\lambda - f^*_\lambda \partial_t f_\lambda\right) = 1, \quad f_\lambda = \frac{e^{-i \sqrt{\lambda^2+m^2} t}}{\sqrt{2\sqrt{\lambda^2+m^2}}} \label{eq:vachar},
\end{gather}
where we have chosen the single-exponent harmonics to find the vacuum solution. Selecting a solution involving two harmonics leads to a state that is a Bogolyubov transformation of the standard vacuum state. For simplicity, we shall restrict ourselves to this specific set of states, though this approach can easily be generalized to cases where the states have a definite number of excitations at each level. Using this decomposition, we can compute all correlation functions that appear when studying the exact correlator \eqref{eq:KStech}
\begin{gather}
  G_{--}(x_1,t_1; x_2 t_2) =  \langle T \phi_i (x_1,t_1) \phi_j (x_2, t_2) \rangle =\notag\\
  =\sum_\lambda \delta_{ij}\left[f_{\lambda}(x_1,t_1) f^*_{\lambda}(x_2,t_2) \theta(t_1- t_2) \right. \notag\\ \left. + f_{\lambda^*}(x_1,t_1) f_{\lambda}(x_2,t_2) \theta(t_2-t_1) \right], \notag\\
  G_{+-}(x_1,t_1; x_2,t_2)= \delta_{ij}\langle \phi (x_1,t_1) \phi (x_2, t_2) \rangle = \notag\\
  =\sum_\lambda f_{\lambda}(x_1,t_1) f^*_{\lambda}(x_2,t_2),  \ldots \label{eq:ansatz},
\end{gather}
where the subscripts $\pm$ indicate the origin of the operators in the correlator \eqref{eq:KStech}
: $+$ for the anti-time-ordered component and $-$ for the time-ordered component. We now study the large
$N$ limit, where careful consideration of all diagrams reveals that only one type of diagram contributes in the large
$N$ limit (the so-called "snail" diagrams \cite{Klebanov:2018fzb}, see Fig. \eqref{fig:dom}).
This results in the following exact Dyson-Schwinger equation:
\begin{gather}
  G_{\sigma_1 \sigma_2}^{-1} = G_{0,\sigma_1\sigma_2}^{-1} -  \delta_{\sigma_1 \sigma_2} \Sigma(x,t), \notag\\
  \Sigma(x,t) = G_{++}(x,t; x,t) = \sigma(x,t),
\end{gather}
where we have accounted for the fact that all propagators coincide at identical points. Plugging the decomposition \eqref{eq:ansatz}
back into this equation, the solution remains formally valid, but we must adjust the equation for the harmonic:
\begin{gather}
  \left(\Box + m^2_0 +\sigma\right) f_\lambda =0, \quad \sigma = \frac{g_0 \Lambda}{N} \sum_{\lambda} \left|f_{i\lambda}(x,t)\right|^2, \label{eq:naiveeq}
\end{gather}
which gives us a system of equations controlling the dynamics of the large $N$ model. By choosing harmonics $f_\lambda(t)$, we select an initial state and can study the behavior of these equations. In this and the following section, we will focus on the case where all states for all
$i=1,\ldots,N$ are identical.

The motivation for considering such highly excited states is straightforward:
such states would have energy on the order of the thermal free energy at finite temperatures with \(\beta = \mathcal{O}(1)\).
Hence, we would expect these states to be relevant at finite temperatures and capable of thermalizing.
Indeed, if we choose an initial state with \(\sigma = \mathcal{O}(N^{-1})\) in the limit \(N \to \infty\), we obtain a free field theory,
which, of course, would not thermalize. However, if the energy of a state is on the order \(E_{\text{st}} \sim N\),
we would expect \(\sigma = \mathcal{O}(1)\), thus causing backreaction on the harmonics and leading to subsequent thermalization.

\section{Real-time equations for the large $N$ vector model}
\label{sec2}
Due to the UV divergences, the equations \eqref{eq:naiveeq} should be reorganized once the continuum limit is considered,
and remain unchanged when \(d < 2\) (for instance, in the case of large \(N\) quantum mechanics \cite{Trunin:2021lwg}).
The first problem arises from UV divergences: the equation for the field \(\sigma\) contains divergences.
To make it finite and thus precise, we restrict the summation to some UV cut-off \(\Lambda\) and subtract a constant from \(m_0^2\)
to compensate for these large contributions. In the end, we obtain a physical mass \(m \ll \Lambda\) and a physical field \(\sigma\).
After implementing these procedures, we arrive at the following equations:
\begin{gather}
  \left(\Box + m^2 +\sigma\right) f_\lambda =0, \notag\\
  \sigma = \frac{g_0 \Lambda}{N}  \sum_{i,\lambda \leq \Lambda} \left(\left|f_{i,\lambda}(x,t)\right|^2-\frac{1}{2\sqrt{\lambda^2+m^2}}\right), \label{eq:dynamics}
\end{gather}
where we have used the harmonics \eqref{eq:vachar} to fix  $m_0^2$ and renormalize \(\sigma\).
These equations are still not entirely correct in the continuum limit \(\Lambda \to \infty\).
The reason is that \(g_0 \Lambda\) is dimensionful and of the order \(\Lambda\).
Since \(\sigma\) determines the mass of our fields, and we want \(\sigma \sim m \ll \Lambda\),
we should impose the following constraint:
\begin{gather}
  \sum_{\lambda \leq \Lambda} \left(\left|f_\lambda(x,t)\right|^2-\frac{1}{2\sqrt{\lambda^2+m^2}}\right) \propto \frac{\sigma}{\Lambda} \to 0\,.\label{eq:con}
\end{gather}
Thus, the role of \(\sigma\) in the above equation is to ensure the constraint \eqref{eq:con}
during the dynamics of harmonics determined by equation \eqref{eq:dynamics}.
These features make the situation quite drastically different from the large \(N\) quantum mechanics.
Consequently, we arrive at the following equations (where we have absorbed \(m^2\) into \(\sigma\)):
\begin{gather}
  \left(\Box + \sigma + m^2\right) f_\lambda =0, \notag\\ \quad \sum_{\lambda\leq \Lambda} \left|f_\lambda(x,t)\right|^2 =  r= \sum_{\lambda \leq \Lambda} \frac{1}{2\sqrt{\lambda^2+m^2}} \label{eq:dynharm}
\end{gather}
Using the constraint for the harmonics $f_\lambda$, we can find  $\sigma$ as a function of harmonics
\begin{gather}
  \sigma = \frac{\sum_\lambda |\partial_\mu f_\lambda|^2}{\sum_\lambda |f_\lambda|^2}  = \frac{\sum_\lambda \left( |\partial_t f_\lambda|^2 - \lambda^2 |f_\lambda|^2\right)}{r} - m^2 \label{eq:sigfund}
\end{gather}
Roughly speaking, we find that the quantum dynamics of the large \(N\) vector model corresponds to the dynamics of a particle on an \(S^{2N_\Lambda-1}\) sphere
of radius \(r\) in a complex \(N_\Lambda\)-dimensional space, where \(N_\Lambda\) is the number of modes with eigenvalues less than the cut-off \(\Lambda\).
Note that such dynamics appear only when the number of excited degrees of freedom is on the order of \(N\); if there were fewer excited degrees of freedom, our constraint would be negligible in the large \(N\) limit, resulting in weakly coupled dynamics. Analogous equations were studied in \cite{Das:2012mt}.

Now, we will restrict ourselves to the case where the field \(\sigma\) is homogeneous. This assumption allows us to analytically discuss the properties of the dynamics, leading to the following equations:
\begin{gather}
  \left(\partial_t^2 + \lambda^2 + \sigma \right) f_\lambda  =0, \label{eq:finaleq}
\end{gather}
where for simplicity we reabsorbed mass $m^2$ into the definition of the field $\sigma$
(we can always split them back). We multiply the above equation by $\partial_t f^*_\lambda$ and taking the real part we get
\begin{gather}
  \partial_t \left[\left|\partial_t f_\lambda\right|^2 + \lambda^2 \left|f_\lambda\right|^2\right] + \sigma \partial_t \left|f_\lambda\right|^2 = 0, \label{eq:energyfunctional}
\end{gather}
if $\sigma=0$ we would have conservation of energy for each individual harmonics,
and this way we could establish the conservation of higher spin currents.
But if we have some non-trivial behaviour of $\sigma$, each individual energy would depend
on time and we would have a violation of conservation of higher spin currents.
Nonetheless, summing over $\lambda$ and using \eqref{eq:con} we get that the following quantity is conserving
\begin{gather}
  E =\frac12 \sum_\lambda \left(\left|\partial_t f_\lambda\right|^2 + \lambda^2 \left|f_\lambda\right|^2 - \lambda\right).
\end{gather}
This is, obviously, the energy of our model, and we have also subtracted the ground state energy for each harmonic \(\lambda\) by setting the energy of the vacuum state to zero.

Looking at equation \eqref{eq:finaleq}, we see that we can treat the harmonics \(f_\lambda\) as fields oscillating in the background of a time-dependent external field \(\sigma\), which is created by other harmonics. This time-dependent external field implies that, for each individual harmonic, the conservation of energy is violated, allowing for particle creation.

Let us assume that, after some time, \(\sigma \to \text{const}\). We then expect the harmonics to behave as
\begin{gather}
  f_\lambda(t) \to \left[\frac{\alpha_\lambda }{\sqrt{2\omega_\lambda}} e^{-i \omega_\lambda t} + \frac{\beta_\lambda }{\sqrt{2\omega_\lambda}} e^{i \omega_\lambda t} \right], \notag\\ \left|\alpha_\lambda\right|^2 - \left|\beta_\lambda\right|^2 = 1,  \quad
  \omega_\lambda^2  = \lambda^2 + \sigma
\end{gather}
the appearance of a sum of two exponents is not something special ---
in the presence of the "external" time-dependent potential we expect particle creation.
And this particle creation is encoded in these coefficients $\alpha_\lambda,\beta_\lambda$.
By doing the Bogoliubov transformation we can find, for instance, the particle number
at level $\lambda$ will be $\left|\beta_\lambda\right|^2$. Now, if we plug definition of $\sigma$ in these equations, we get
\begin{gather}
  \sum_\lambda |f_\lambda|^2 = \sum_{\lambda \leq \Lambda} \frac{1}{2\sqrt{\lambda^2+\sigma}}\left(\left|\alpha_\lambda\right|^2 + \left|\beta_\lambda\right|^2 + \right. + \notag\\+ \left. \alpha\beta^* e^{2i\omega_\lambda t} + \alpha^*\beta e^{-2i\omega_\lambda t} \right)
\end{gather}
In the limit $t\to \infty$ we can neglect the oscillating terms
\footnote{That could be justified in the continuum limit $\Lambda \to \infty$.} and get the following equation on $\sigma$
\begin{gather}
  \sum_\lambda \left(\frac{1+2\left|\beta_\lambda\right|^2}{2\sqrt{\lambda^2+\sigma}} - \frac{1}{2\lambda}\right)=0, \notag\\
  \sum_\lambda \frac{\left|\beta_\lambda\right|^2}{\sqrt{\lambda^2 + \sigma}}= \sum_\lambda \frac{\sigma}{2\lambda\sqrt{\lambda^2+\sigma} (\sqrt{\lambda^2+\sigma} + \lambda)}
\end{gather}
that gives the gap equation for $\sigma$ as a function of $\left|\beta_\lambda\right|^2$. If we compute now the energy we get
\begin{gather}
  E = \sum_\lambda \left[\frac{(1+2\left|\beta_\lambda\right|^2) (2\lambda^2+\sigma)}{4\sqrt{\lambda^2+\sigma}} - \frac{\lambda}{2}\right] = \\
  =\sum_\lambda\left[  \left|\beta_\lambda\right|^2 \sqrt{\lambda^2+\sigma} + \frac12 \sqrt{\lambda^2+\sigma} - \frac{\sigma}{4\lambda}-\frac{\lambda}{2}\right], \notag
\end{gather}
where we have used the gap equation. The meaning of this equation is quite simple,
the first part says that we have $\left|\beta_\lambda\right|^2$ particles at level $\lambda$ with energy $\sqrt{\lambda^2+\sigma}$
and the second part just computes the offset of the ground state energy, caused by the change of the mass by the field $\sigma$.

We can generalize the above the equations when we have a pure state
with $n_\lambda$ excitations (that corresponds to just the change $\left|\beta_\lambda\right|^2 \to n_\lambda$)
at each energy level $\lambda$. Thus we can find that if we have a set of states $n_\lambda$
and a field $\sigma$ that satisfy the following equation
\begin{gather}
  \sum_\lambda \left(\frac{1+2 n_\lambda}{2\sqrt{\lambda^2+\sigma}} - \frac{1}{2\lambda}\right)=0,
\end{gather}
then one can find that we will have a state of our field with energy
\begin{gather}
  E =N \sum_\lambda\left[  n_\lambda \sqrt{\lambda^2+\sigma} + \frac12 \sqrt{\lambda^2+\sigma} - \frac{\sigma}{4\lambda}-\frac{\lambda}{2}\right], \label{eq:Esig}
\end{gather}
let us note the above expression satisfies the following relation
\begin{gather}
  \frac{\partial E(\sigma)}{\partial \sigma} = N \sum_\lambda \left[\frac{1+2 n_\lambda}{4\sqrt{\lambda^2+\sigma}} - \frac{1}{4\lambda} \right] =0,
\end{gather}
so for given $n_\lambda$ we can find that energy by minimizing the functional \eqref{eq:Esig} as a function of $\sigma$.

Now, we want to address the question of thermalization. We can study the equations \eqref{eq:dynharm} numerically. For that, we will set \(\mathcal{M} = S^2\) and \(m^2 = \frac{1}{4}\), which gives the spectrum \(\lambda = L + \frac{1}{2}\), where \(L \in \mathbb{N}_{\geq 0}\), corresponding to the critical \(O(N)\) model.

\begin{figure}
  \centering
  \includegraphics[width=\linewidth]{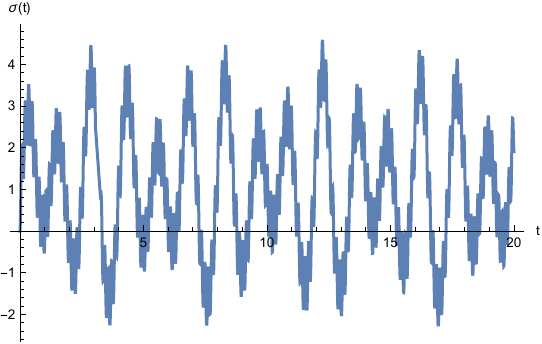}
  \caption{The plot of $\sigma$ as a function of time $t$ of $m^2=\frac14$ and homogeneous initial perturbation.}
  \label{fig:sig(t)pdf}
\end{figure}

After this setup, we can either work directly with equations \eqref{eq:dynharm} and \eqref{eq:sigfund}, or we can use \eqref{eq:naiveeq} and choose a large coupling constant. The results will be equivalent, but the latter approach tends to produce more stable results. We observe that the resulting dynamics does not stabilize and thus does not thermalize (see Fig. \eqref{fig:sig(t)pdf}).

The reason is that we have a large number of commuting charges \cite{deift1980nonlinear} for any choice of the spatial manifold \(\mathcal{M}_2\) and physical mass \(m\).
For instance, we can introduce
\begin{gather}
  I_\lambda = \left|f_\lambda\right|^2 + \sum_{\mu \neq \lambda}\frac{(f_\lambda \partial_t f^*_\mu - f^*_\mu \partial_t f_\lambda)^2}{\lambda^2 - \mu^2},
\end{gather}
which can easily be checked to satisfy \(I'_\lambda(t) = 0\). Thus, the question remains open as to whether the large \(N\) vector model does indeed thermalize.
Since these charges commute, they do not form the higher spin algebra \cite{Vasiliev:1990en}. Note that these charges appear only in the case of a homogeneous perturbation. In the case of an inhomogeneous perturbation, the resulting dynamics should not be integrable, and we would expect the background \(\sigma\) field to induce particle production. Additionally, one possible way to achieve thermalization is to consider the system of equations in the continuum limit (taking \(\Lambda \to \infty\)) and show that in this particular limit, we could still obtain solutions that thermalize (or at least where \(\sigma(t)\) tends to a stationary value) by dispersing energy to higher energy modes.

\section{Computation of the thermal free energy}
\label{sec3}
In the previous sections, we restricted ourselves to the case where we have a state corresponding to a Bogolyubov transformation of a standard Fock vacuum. Here, we will consider the case of usual Fock states with definite number of excitations at energy level and find the corresponding spectrum of these states. Thus, we consider states of the form
\begin{gather}
  \ket{n_{i\lambda}} = \prod_\lambda \prod_i a^\dagger_{i\lambda} \ket{0}, \quad f_\lambda(t) = \frac{e^{-i\omega_\lambda t}}{\sqrt{2\omega_\lambda}}
\end{gather}
where we have a macroscopically large number of operators \(a^\dagger\). By studying the dynamics of such states using the Keldysh-Schwinger technique, we find that these states are stationary, with \(\sigma = \text{const}\) as a homogeneous background. Thus, \(\ket{n_{i\lambda}}\) is an eigenstate of this system, and we can use equation \eqref{eq:Esig} to determine the energy of these states.  Note that if we constrain to the case where only \(n_{i, \lambda = 0}\) are non-zero, we recover the result obtained by Giombi et al. \cite{Giombi:2020enj}, where the energies or dimensions of such states were computed using different methods.

To do so, we introduce the averaged number of states with eigenvalue \(\lambda\) as
\begin{gather}
  n_\lambda = \frac{1}{N} \sum^N_{i=1} n_{i\lambda}, \notag
\end{gather}
where the macroscopically large number of particles implies that these \(n_\lambda\) are non-zero. Then, we find that the energy of such a state is given by
\begin{gather}
  E(n_\lambda;\sigma) =N \sum_\lambda\left[  n_\lambda \sqrt{\lambda^2+\sigma} + \frac{1}{2} \sqrt{\lambda^2+\sigma} - \frac{\sigma}{4\lambda}-\frac{\lambda}{2}\right], \notag\\
  \quad \frac{\partial E(\sigma)}{\partial \sigma} = 0,
\end{gather}
where the last equation determines \(\sigma_{n_\lambda}\) as a function of \(n_\lambda\).

How many states have \(n_\lambda\) particles at level \(\lambda\)? This requires distributing \(N n_\lambda\) excitations over \(N\) oscillators.  This is a simple combinatorics problem, and we can compute the entropy contribution of such a configuration:
\begin{gather}
  S_{n_\lambda} = \log \binom{N}{N(1+n_\lambda)} \approx \notag\\
  \approx N \left[(1+n_\lambda) \log\left(1+n_\lambda\right) - n_\lambda \log n_\lambda \right].
\end{gather}
Then, our partition function is given by
\begin{gather}
  Z = \int \prod_\lambda dn_\lambda e^{S_{n_\lambda} - \beta E(n_\lambda;\sigma_{n_\lambda}) }.
\end{gather}
Since energy and entropy scale with \(N\), we can use the saddle-point approximation, yielding
\begin{gather}
  \frac{\partial}{\partial n_\lambda}\left[  S_{n_\lambda} - \beta E(n_\lambda;\sigma_{n_\lambda})\right] = \log\left( 1+ \frac{1}{n_\lambda}\right) - \beta \sqrt{\lambda^2+\sigma}, \notag\\ \quad n_\lambda = \frac{1}{e^{\beta \sqrt{\lambda^2 + \sigma}} - 1}.
\end{gather}
The equation for \(\beta\) is
\begin{gather}
  \sum_\lambda \left[ \frac{\coth \left(\frac{\beta}{2}\sqrt{\lambda^2+\sigma} \right)}{\sqrt{\lambda^2+\sigma}} - \frac{1}{\lambda}\right] = 0,
\end{gather}
which allows us to find a typical state contributing to this energy. We thus find that
\begin{gather}
  \ket{\Psi} = \prod_\lambda \prod_{k=1}^{n_\lambda} a^\dagger_{i_k \lambda} \ket{0}, \quad n_\lambda = \frac{1}{e^{\beta\sqrt{\lambda^2+\sigma}} -1}.
\end{gather}
Restricting ourselves to the case of \(S^2\) and setting \(m=\frac{1}{4}\), we find that in the limit \(\beta \to 0\), the thermal free energy of the critical \(O(N)\) model is
\begin{gather}
  F_{\rm crit} = E - T S \approx  -\frac{2\zeta(3) N}{5\pi \beta^3},
\end{gather}
which coincides with the free energy of the critical \(O(N)\) model on a plane computed by conventional methods \cite{chubukov1994theory}. Using the operator-state correspondence, we find that the typical thermal state should be described by the operator
\begin{gather}
  O_\beta = \prod_{k=1}^{n_0} \phi_{i_k} \prod_{k=1}^{n_1} \partial_{\mu^1_k}\phi_{i_k} \prod_{k=1}^{n_2} \partial_{\mu^1_k} \partial_{\mu^2_k }\phi_{i_k} \ldots
\end{gather}
Hence, we expect that if we compute the thermal one-point functions of various operators \(\mathcal{O}\), they will coincide with the OPE coefficients \(C_{\beta \beta \mathcal{O}}\).

\section{Conclusion}
The large \(N\) models offer an ideal framework for addressing various conceptual and technical questions in different areas of quantum field theory.
In our study of the real-time dynamics of the large \(N\) model, we demonstrated that at energies \(E = \mathcal{O}(N)\), the dynamics of the model becomes strongly coupled but is still governed by a simple set of equations. In the case of homogeneous perturbations, we showed that the system becomes integrable. A detailed exploration of this model would be interesting, particularly to determine whether the large \(N\) vector models thermalize in the continuum limit. Additionally, it would be interesting to apply the developed formalism to study the dynamics of quantum field theories in the presence of external time-dependent fields and to derive the hydrodynamic description of the emergent IR dynamics.

\section*{Acknowledgements}
We thank Igor Klebanov, Simone Giombi, Zohar Komargodski,  Yifan Wang, Gabriel Cuomo and Alexei Milekhin for helpful discussions and comments on the draft.
\bibliography{defect}
\end{document}